\documentclass{article}

\input{tcilatex}
\begin{document}

\begin{center}
\bigskip

{\Huge An Alternative Explanation for Cosmological Redshift}

\bigskip

{\Large David Schuster\footnote{%
dschuste@mines.edu}}

\textit{Department of Physics, Colorado School of Mines}

\bigskip

\bigskip

{\Large Abstract}
\end{center}

\qquad

\qquad The first and most compelling evidence of the universe's expansion
was, and continues to be, the observed redshift of spectra from distant
objects. \ This paper plays "devil's advocate" by providing an alternative
explanation with elementary physics. \ I\ assume a steady-state universe
that is infinite in both expanse and age, with the observed redshifts caused
by particle interactions creating an overall index of refraction of the
universe. \ The cumulative effects of these interactions over long distances
cause not only the shifts that we observe, but also the monotonically
increasing redshifts as more distant objects are observed. \ This is a novel
explanation for the phenomenon known as "tired light" which has been
discussed for decades.

\bigskip

\begin{center}
{\Large 1. Introduction}

\bigskip
\end{center}

Hubble was the first to describe the relation between distance from Earth
and radial velocity of extra-galactic targets in 1929 [1]. \ This
ground-breaking discovery has served as a foundation on which to build all
of modern Big Bang cosmology. \ Hubble's work is based on the apparent
Doppler redshift of spectra obtained from these targets. \ As is widely
known to any physics student, a Doppler redshift occurs when an emitter is
moving away from an observer in the observer's frame. \ Redshift is given
simply by:

\begin{equation}
z=\frac{\lambda _{observed}-\lambda _{emitted}}{\lambda _{emitted}}
\end{equation}

Where $U\ll $ $c$ and is the velocity of the emitter. \ This concept is
applied cosmologically to calculate objects' radial recession velocities as
observed from Earth. \ It is the contention of this paper that an
alternative explanation for this effect can be offered by examining a flaw
in the basic assumptions made about the nature of the observations. \ The
flaw is that the intervening space between the emitter and the observer is
perfect vacuum, allowing the photons to travel at exactly $c$. \ The ISM and
IGM have some mass density and, while it may be extremely low, it cannot be
ignored when examining optical sources in space. \ This mass density creates
an index of refraction that slows the light down and causes an apparent
redshift.

\bigskip

\begin{center}
{\Large 2. Putting the Brakes on Light}

\bigskip
\end{center}

Einstein set the universal speed limit at $c$ with Relativity, he also
claimed that light always traveled at $c$ in vacuum, regardless of reference
frame.{\Large \ \ }However, in materials, it is well established that light
will slow down or even stop, depending on the conditions [2]. \ To a physics
student, this is why the image of a drinking straw appears broken in a glass
of water, the light from the image is being redirected as it slows down
through the medium. \ The quantitative description of this is the index of
refraction given by:

\begin{equation}
n=\frac{c}{v}
\end{equation}

Where $v$ is velocity of light as is propagates through the material. \ In
vacuum, $c$ is related to frequency and wavelength by $c=f\lambda $. \ In a
material, however, the frequency stays constant, but the wavelength shifts
according to the change in velocity implying that $n_{1}\lambda
_{1}=n_{2}\lambda _{2}$ when a light beam goes from a material with index $%
n_{1}$ to a material with index $n_{2}$. \ This effect is traditionally
compared to a marching band that goes from marching on dry hard ground to
soft wet mud; the lines of the band will remain oriented the same relative
to one another within the two media, however the lines will be redirected as
the marchers slow down in the mud. \ 

\bigskip

\begin{center}
{\Large 3. The IGM\ as a Super-Low Density Fluid}

\bigskip
\end{center}

Current models of the intergalactic medium contend that it has mass density
on the order of $\rho _{M}\approx 10^{-27}\frac{kg}{m^{3}}$. \ While it is
true that this equates to approximately one atom of neutral Hydrogen per
cubic meter, averaging over cosmological distances, it is reasonable to
consider the IGM a super-low density fluid. \ Modeling the IGM in this
manner does away with the idea that light propagates exactly at $c$ through
intergalactic space. \ Instead the IGM now has an index of refraction:

\bigskip

\begin{equation}
n_{IGM}=1+\epsilon
\end{equation}

\bigskip

In this classical framework, light certainly has a wavelength shift that
deviates from the wavelength of light in a perfect vacuum. \ Given that $%
n_{1}\lambda _{1}=n_{2}\lambda _{2}$, if our reference wavelength $\lambda
_{c}$ occurs when $n_{2}=1$, corresponding to a perfect vacuum, then:

\bigskip

\begin{equation}
(1+\epsilon )\lambda _{emitted}=\lambda _{c}
\end{equation}

\bigskip

Which then implies that the redshift measured is given by:

\bigskip

\begin{equation}
z=\frac{\lambda _{observed}-\lambda _{emitted}}{\lambda _{emitted}}=\frac{%
\lambda _{c}-\frac{\lambda _{c}}{1+\epsilon }}{\frac{\lambda _{c}}{%
1+\epsilon }}=\epsilon
\end{equation}

\bigskip

So far we have played some games with elementary physics by coming up with
an expected \textit{constant} redshift. \ While these games may have been
fun, the idea of a constant redshift has already been disproven by mountains
of observational evidence showing that redshift gets larger and larger the
farther away we look. \ To resolve this we must go further and examine the
behavior of the photon as it travels to an observer from a distant target.

\bigskip

\begin{center}
{\Large 4. Variable Optical Travel Time}

\bigskip
\end{center}

Let us perform a short thought experiment. \ Imagine two light-emitting
objects an equal distance $x$ away from an observer. \ They each emit a
photon at the same time toward the observer, the difference is that the path
between the first object and the observer is perfect vacuum with $n=1$ and
the path between the second object and the observer has the particle density
of the IGM with $n=1+\epsilon $. \ Assuming that the observer and the
sources share a common frame, then the travel time for the photon from
object number one is $t_{1}=\frac{x}{c}$. \ The travel time for the second
object's photon will be delayed, however by an amount proportional to the
number $N$ of absorptions and reemissions experienced in the intervening
medium. \ If the characteristic, average time delay from these reactions is $%
t_{0}$ then the travel time for object two's photon is:

\bigskip

\begin{equation}
t_{2}=t_{1}+Nt_{0}=\frac{x}{c}+Nt_{0}
\end{equation}

\bigskip

Obviously, as the density of the intervening medium increases, so does the
number of interactions and, consequently, so does the travel time of the
light. \ This is the effect seen in a dense material like calcite where
there are so many interactions that the light slows down appreciably in a
short distance. \ Thus, as the path $x$ gets longer and more interactions
occur, travel time increases. \ This gives the photon an effective velocity
of:

\bigskip

\begin{equation}
v=\frac{x}{t_{1}+Nt_{0}}
\end{equation}

\bigskip

With the index of refraction given by:

\bigskip

\begin{equation}
n=1+\epsilon =\frac{c}{v}=\frac{c(t_{1}+Nt_{0})}{x}
\end{equation}

\bigskip

Recalling that $\epsilon $ is our value for redshift, simple algebra will
give us a variable value for redshift of the form:

\bigskip

\begin{equation}
z=N\frac{t_{0}}{t_{1}}
\end{equation}

\bigskip

Obviously, as we increase the distance to a source, both $N$ and $t_{1}$
will increase. \ The traditional model assumes a more-or-less linear
relationship between distance (or photon travel time) and redshift. \ This
model differs in that larger values of $t_{1}$ actually work to suppress
redshift, whereas $N$ works to increase redshift. \ Thus, to fit current
observations, $N$ must grow faster relative to $t_{1}$.

A "back-of-the-envelope" calculation for the mean free path of a photon at
an average number density of $n=1\frac{particle}{m^{3}}$ gives:

\bigskip

\begin{equation}
\ell =\frac{1}{n\sigma }\approx 37.6kly
\end{equation}

\bigskip

Assuming the interaction cross-section to correspond to the Bohr radius. \
This means that a photon will, on average, have an interaction and,
accordingly, a characteristic delay every $37600$ light years. \ This is
using the minimum particle density in intergalactic space, which can vary
widely up to approximately $1000\frac{particles}{m^{3}}$ in areas of
particularly high density. \ This fact turns out to be very fortunate for
our analysis, however. \ If the particle density remained constant at this
minimum value for the duration of the photon's journey, then the growth rate
of the number of interactions $N$ and the growth of the unperturbed
travel-time $t_{1}$ would offset one another and we could expect a flat
redshift curve everywhere we looked. \ This is clearly not the case
observationally. \ This fact is why it is important to consider the total
number of interactions, rather than attempting to average over large
distances since, in our model, the universe is a fluid of continuously
varying density. \ The effect must be considered cumulatively, allowing the
interaction delays to increase multiplicatively through regions of high and
low density. \ The regions of high density, relative to those of the low
density, are what will work to increase $Nt_{0}$ compared to $t_{1}$ and,
since photons from more distant objects will statistically travel through
more "stuff" of varied density, observed redshifts for more distant objects
will be higher.

\bigskip

\begin{center}
{\Large 5. Possible Criticism}

\bigskip
\end{center}

There is ample opportunity to criticize this formulation and I'm sure the
knowledgable reader can come up with many more criticisms (after all isn't
that what science is all about?), but for the sake of this paper, I will
address three main concerns:

\bigskip

\begin{center}
\textbf{Why discount Doppler redshift altogether?}
\end{center}

\bigskip

While the way in which the idea was formulated, it may appear that I\ reject
the idea that cosmological objects can have any relative motion. \ This is
not true, as I\ believe that galaxies do move relative to one another,
sometimes blueshifted, sometimes redshifted. \ This formulation was done in
a highly simplistic toy universe simply for the purpose of exploring the
effect of intergalatic particles on photons as they propagate through space
and how it might affect redshift. \ What I\ wanted to address was why we see
redshift everywhere we look and it is my belief that the effect described in
this paper overwhelms any actual relative motion between objects.

\bigskip

\begin{center}
\textbf{Why do we not see large redshifts for local stars known to be
embedded in dust?}

\bigskip
\end{center}

According to my argument, it would make sense that any light emitting object
embedded in particles would automatically exhibit a high redshift. \ This is
not necessarily true, however, since it is only the total number of particle
interactions that correspond to redshift, not the density. \ While the
density of the intervening material certainly plays a part, it is not the
whole story. \ My idea changes the yardstick by which we measure
cosmological distance and a high-redshift quasar could be a great deal
further away than we currently believe it to be. \ This means that a photon
emitted by that quasar could, in theory, have many orders of magnitude more
interactions than a photon emitted from a local star embedded in a dense
dust cloud. \ 

\bigskip

\begin{center}
\textbf{Isn't this just a retread of the Wolf Effect? }[4]

\bigskip
\end{center}

\bigskip While on the surface, my formulation may seem similar to that of
Emil Wolf, I believe that the mechanism by which the redshift is realized is
unique.

\bigskip

\begin{center}
\ {\Large 6. Conclusion}

\bigskip
\end{center}

This model describes an alternate explanation for cosmological redshift and
the supposed relationship between radial velocity and distance. \ The
wavelength shifts observed are postulated to occur not from a Doppler
effect, but rather from an overall, variable index of refraction for an
infinite steady-state universe. \ The time delay from interactions with
particles in the IGM account for the "tired light"\ aspect of this model, as
well as the redshift. \ The variability of the IGM's particle density
accounts for variability of the refractive index and, consequently, the
increase of redshift value with respect to distance. \ The full consequences
and additional ideas for such a "steady-state revival" will be discussed at
length in future papers.

\bigskip

\begin{center}
{\Large 7. References}

\bigskip
\end{center}

[1] Hubble, Edwin, "A Relation between Distance and Radial Velocity among
Extra-Galactic Nebulae", Proceedings of the National Academy of Sciences of
the United States of America, Volume 15, Issue 3, pp. 168-173 (1929)

\bigskip

[2] Yanik M.F., Fan S., "Stopping Light All Optically", Physical Review
Letters, vol. 92, Issue 8, id. 083901 (2004)

\bigskip

[3] Crawford, David F., "Curvature pressure in a cosmology with a
tired-light redshift", Australian Journal of Physics, vol. 52, pp. 753-777
(1999)

\bigskip

[4] Wolf, Emil, "Noncosmological redshifts of spectral lines", Nature 326:
pp. 363-365 (1987)\ \ 

\ {\Large \ }

\begin{center}
\bigskip
\end{center}

\bigskip

\bigskip

\ \ 

\bigskip

\ 

\end{document}